# Evaluating and comparing gender bias across four text-to-image models

**Zoya Hammad[1], Nii Longdon Sowah[2]**

[1]High School, International Community School of Abidjan, Abidjan, Ivory Coast
[2]Computer Engineering Department, University of Ghana, Accra, Ghana

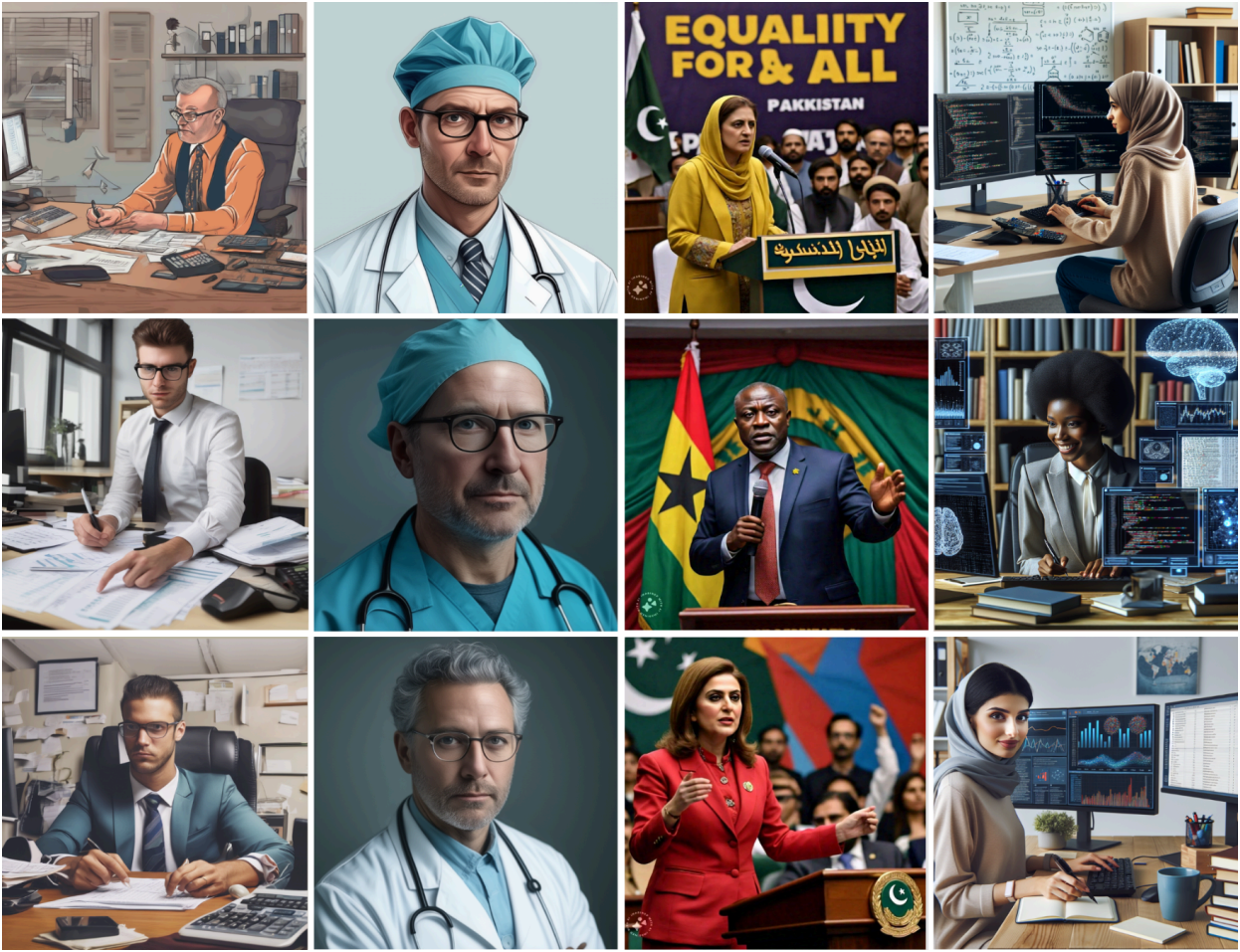

Stable Cascade
(An Accountant)

Stable Diffusion XL
(A Surgeon)

Emu
(A Politician)

DALL-E 3
(A Data Scientist)

**Figure 1. AI-generated Images.** Left to right: Stable Cascade (an accountant - all male), Stable Diffusion XL (a surgeon - all male), Emu (a politician - diverse results), DALL-E 3 (a data scientist - all female).

**SUMMARY**

As we increasingly use Artificial Intelligence (AI) in decision-making for industries like healthcare, finance, e-commerce, and even entertainment, it is crucial to also reflect on the ethical aspects of AI, for example the inclusivity and fairness of the information it provides. In this work, we aimed to evaluate different text-to-image AI models and compare the degree of gender bias they present. The evaluated models were Stable Diffusion XL (SDXL), Stable Diffusion Cascade (SC), DALL-E and Emu. We hypothesized that DALL-E and Stable Diffusion, which are comparatively older models, would exhibit a noticeable degree of gender bias towards men, while Emu, which was recently released by Meta AI, would have more balanced results. As hypothesized, we found that both Stable Diffusion models exhibit a noticeable degree of gender bias while Emu demonstrated more balanced results (i.e less gender bias). However, interestingly, Open AI's DALL-E exhibited almost opposite results, such that the ratio of women to men was significantly higher in most cases tested. Here, although we still observed a bias, the bias favored females over males. This bias may be explained by the fact that OpenAI changed the prompts at its backend, as observed during our experiment. We also observed that Emu from Meta AI utilized user information while generating images via WhatsApp. We also proposed some potential solutions to avoid such biases, including ensuring diversity across AI research teams and having diverse datasets.

## INTRODUCTION

Artificial Intelligence (AI) has been growing remarkably in recent years, impacting numerous aspects of our daily lives. One such area of significant advancement is text-to-image generation. Models like DALL-E and Stable Diffusion have revolutionized the way we create visuals by allowing users to generate images based solely on textual descriptions (1,2). This technology holds immense creative potential. In fact, companies like Levi's are already using it for advertising (3). However, concerns have emerged regarding potential biases within these AI systems (4, 5). In a study conducted by Charles Sturt University, OpenAI's DALL-E was prompted to generate images of medical students and pharmacists. Although 64% of pharmacists in Australia are female, DALL-E failed to reflect this, with only 29.7% of the generated images representing women (6). A similar study conducted at Stanford University also depicted how DALL-E and Stable Diffusion reinforce gender stereotypes when prompted to generate images of occupations or traits (7). For example, software engineers were shown to be male while housekeepers were shown to be female (7). Other studies that evaluated gender bias in Stable Diffusion only also highlighted gender biases such that the aircraft pilots were always male while dieticians were always female (8, 9).

Considering that such AI tools have become easily accessible across all age groups, inherent biases can have significant negative consequences as they can influence one's self-esteem and identity (10). The stereotypes an AI tool may have inherited can severely impact marginalized groups such as women aspiring to become software engineers.

Up until this paper manuscript was drafted, the pPrevious research on gender equity in AI has mostly focused on only one image generation model (e.g., DALL-E or Stable Diffusion) or at max two models (DALL-E and Stable Diffusion) (6-13). Emu, which is a relatively new model, has not been studied or compared before. Additionally, the results from previous studies were not always quantified in a comparative nature. For example, in the study conducted at Stanford University (7), the gender biased outcomes within the two models were only textually pointed out. Since sufficient numerical data was not provided, the levels of bias could not be compared. Additionally, there was no comparison between the two used models and hence, there was no quantifiable distinction between results from Stable Diffusion and DALL-E. Other relevant research, for example the one by Charles Sturt University only focused on a specific profession (e.g, medical professions within DALL-E) (6).

In this research project, we investigated and compared the prevalence of gender bias in AI image generation across four models and 30 professions. We focused on four text-to-image models, DALL-E 3 (1) (14), Emu which is recently released by Meta AI (15), Stable Diffusion XL (16), and Stable Cascade (17), to explore how they represent various professions across genders and whether there has been an improvement in these models over time in terms of navigating through biases. While DALL-E and Stable Diffusion have already been explored to a certain extent, an in-depth comparison with Emu has not been conducted yet. We also used techniques like binomial distribution to evaluate the degree of bias exhibited by different models for different professions.

We focused on the ratio between men and women and detected an imbalance if it deviated significantly from a 50:50 ratio as the first indication of unwanted bias. Here, we focused on occupational gender bias, where bias refers to the tendency of AI models to produce results that reflect occupational stereotypes about a specific gender. By analyzing the generated images, we aimed to identify any patterns of gender imbalance and possibly gender bias, and to understand the potential impact they would have on our perception of career opportunities (**Figure 1**). When AI image generators consistently reinforce stereotypical associations between professions and gender, they can limit our perception of who can succeed in those fields (18, 19). For instance, if an AI consistently portrays doctors as male and nurses as female, it perpetuates existing gender imbalances within the healthcare sector.

This research project compares DALL-E, Emu, Stable Diffusion XL (SDXL) and Stable Cascade (SC) in terms of their gender representation in profession-based image generation. Based on existing research, we hypothesized that all tools would exhibit some degree of gender bias (6-9). However, since Emu is the most recent development, released after several critical discussions around fairness in AI (6, 7, 8, 9), we expected that the developers would have taken adequate steps to ensure diverse training data and hence, more gender-balanced results. Since Open AI and Meta AI have not publicly released intricate details regarding their training datasets, it is not feasible to compare them with Stable Diffusion models to conclude potential reasons of bias. Regardless, our hypothesis stood correct for Stable Diffusion models as we observed that being comparatively older models, they exhibited a great degree of gender bias, favoring men in higher-education and higher-paying professions. Similarly, our hypothesis for Emu seemed correct as it depicted comparatively gender-balanced results. It seems that after severe backlash (20, 21, 22), companies like MetaAI are actively taking steps to establish ethical guidelines for their development teams.

Interestingly, Open AI's DALL-E exhibited results that were almost opposite to those from Stable Diffusion models, such that the ratio of women to men was almost always significantly higher. Here we still observed a bias, but in this case, the bias was favoring females over males. This could be explained by the fact that OpenAI changes the prompt statements at its backend, as observed during our experiment. Perhaps this is an attempt to "correct" results and avoid controversies (20, 21, 22).

This research holds significance for several reasons. Firstly, it contributes to our understanding of the limitations of AI and the importance of responsible development practices. Secondly, by uncovering gender bias in image generation, we take the first step towards a discussion on who decides on what an appropriate representation (i.e generated gender ratio) is. Does this responsibility fall on the engineers developing such algorithms or should the leading members of AI companies decide on these gender representation ratios? Should public users have a say in this? By initiating such conversations, we can work towards creating fairer and more inclusive AI tools that represent the diversity of the real world.

**RESULTS**

For each of the four image generating models, we generated 50 images for 30 different professions **(Figures 1, 2)**. We first calculated the relative percentages of men and women represented for all four models (**Figure 2**). This gives us a summary of which model favors which gender across different professions. For example, 3 out of 4 models (Emu, Stable Cascade, SDXL) predominantly generated male images for professions such as CEO, pilot, and scientist, reinforcing traditional gender stereotypes in high-status or STEM-related fields. However, OpenAI's DALL-E consistently favored women for such professions.

Individual visualizations for each text-to-image model can be found in **Figure 3** (DALL-E 3), **Figure 4** (Emu), **Figure 5** (Stable Diffusion XL), and **Figure 6** (Stable Cascade). **Figure 7** further demonstrates the probabilities of generating male versus female images for each profession, overall across the four models tested. Through our generated images, observed a clear pattern, where traditionally male-dominated professions such as pilot (75.75% male), CEO (77.50% male), and engineer (77.75% male) are overwhelmingly represented by images of males. Conversely, for professions that are stereotypically associated with women, such as nurse (91.50% female), housekeeper (63.25% female), and administrative assistant (87.50 % female), the models were much more likely to generate images of females.

Furthermore, **Figure 8** presents the p-values obtained from binomial tests assessing gender biases across the four generative AI models for different professions. We conducted the binomial test to determine whether the proportion of images depicting males versus. females generated for each profession significantly deviates from an expected 1:1 ratio (50% male, 50% female). A small p-value (<0.05) suggests a significant deviation from this equal distribution, indicating gender bias in the model's outputs. Conversely, a higher p-value suggests no strong evidence of bias. In the color-coded **Figure 8**, where red represents "statistically significant" while blue represents insignificance, we can observe that more than 110 out the 120 cases (91.6%) exhibit a statistically significant degree of gender bias. Given that there is no clear, distinct pattern of gender bias in the output images of the four models, we can say that none of the four models seem to be performing better than another when it comes to generating images with a 1:1 gender ratio.

It is important to note that 0.0 values in the table are the result of rounding; these p-values are extremely small (i.e., below 0.001) but are not exactly zero. Such small p-values indicate strong statistical significance, suggesting that gender bias is highly unlikely to have occurred by random chance in these cases. Professions such as CEO, CFO, and doctor consistently exhibit these near-zero p-values across all models, reinforcing the presence of systemic biases in AI-generated outputs.

Additionally, **Figure 9** depicts binomial distribution calculations to determine the probability of attaining a 1:1 gender ratio across different models and different professions. As observed through **Figure 9**, the probability of achieving a perfectly balanced ratio is extremely low. Even the highest probability, i.e., for Stable Cascade and "a person cooking in the kitchen" is below 0.10. Hence, it appears that with the current state of these models, achieving a 1:1 gender ratio is highly unlikely. This may change in the upcoming months as companies try to include diversity into algorithms.

We will first describe the results from Stable Diffusion models and Meta AI's Emu . When generating images for different professions, the general pattern across different models remained similar: high-paying professions, including managerial roles such as CEO or CFO, were dominated by images of men. In fact, Stable Diffusion XL and Stable Cascade showed no diversity and generated 100% images with males for these professions **(Figure 5, 6)**. Meta AI generated more images of women, hence depicting at least some diversity **(Figure 4).** Similarly, when generating images of roles that require higher education, for example a chemical engineer, lawyer, or scientist, we again observed that most images generated consisted of male figures. We also observed that some professions, despite being in the same domain, still depicted a great degree of gender bias. For example, when generating images of a doctor, both SDXL and SC returned 100% male images while Emu returned 88% male images. However, when generating images for a nurse, the numbers seemed reversed. Both SDXL and SC returned 100% female images while Emu returned 82% female images. Similarly, when generating images for "a person cooking in the kitchen," a greater percentage of images were of females. However, when using the prompt "a Chef," which implies a more formal profession compared to "a person cooking in the kitchen," the majority of generated images were of men. We also observed this pattern in images of teachers versus images of professors. Even though the two professions fall under the same domain (i.e., education), earning the title of Professor requires a higher level of education. We observed that when generating images of a Professor, the majority of images were of men. On the other hand, when prompting for a Teacher, the AI-generated images mostly consisted of women. This demonstrates a significant amount of gender bias in AI-image generation tools.

When looking at the results from OpenAI's DALL-E 3, 28 out of 30 professions had more female images (**Figure 2, 3**). OpenAI's results were strikingly different from other tested models. For example, for a surgeon, SC, SDXL, and MetaAI's Emu generated 100%, 98%, and 82% male images respectively. On the contrary, OpenAI generated 82% female images, depicting almost opposite results in comparison to the other models. Similarly for a CEO, SC and SDXL generated 100% male images while Emu generated 98% male images. However, OpenAI generated only 22% male images and 78% female images **(Figure 2, 3)**. As observeOpenAI achieves these results by modifying written prompts at the backend and adding certain keywords to ensure diversity. We will discuss this further in the next section.

**DISCUSSION**

In this research, we aimed to explore gender bias across four AI-image generation models. We hypothesized that the comparatively older models, DALL-E and Stable Diffusion, would exhibit a noticeable degree of gender bias favoring men, while Emu, which was recently released by

Meta AI, would have more balanced results. As hypothesized and also observed in previous related studies, Stable Diffusion models exhibited a noticeable degree of gender bias while Emu demonstrated more balanced results (7-9). Interestingly, Open AI's DALL-E exhibited almost opposite results, such that the ratio of women to men is almost always significantly higher. OpenAI's results still exhibited a bias, but in this case, favoring women over men. These results were inconsistent with those from previous studies. For example, when researchers at Charles Sturt University generated images of medical professionals using DALL-E 3, only 29.7% of images represented women (6). However, in our experiment, when we generated images of doctors and surgeons, 82% of the images represented women. It is possible that in the time between the two studies, OpenAI has made an effort to overcome previous reports on bias (6) by modifying their systems. We will discuss this further in the coming paragraphs. Another study at Stanford University that used DALL-E 2 for their experiments also observed a significant occupational bias in their results, such that most generated software engineers were male instead of female (7). Although their paper does not provide any statistics that can be compared with ours, we suspect that improvements have been made to DALL-E 3, eliminating negative bias against women. In fact, it seems to be favoring women instead.

Although we see improvement in DALL-E, our results for Stable Diffusion (SDXL and SC) were consistent with previously conducted experiments. For example, an experiment by Bloomberg Technology pointed out that although the actual percentage of female doctors is 39%, only 7% of the images generated by Stable Diffusion represented women (23). In our experiment, we utilized Stable Diffusion XL and Stable Cascade. Both of these models generated 100% male images. Bloomberg's authors did not mention which version they used for their experiment. While we used the latest version, our results seemed consistent with Bloomberg's findings.

Text-to-image models are trained on large datasets containing millions of images along with their text descriptions. ImageNet is one such renowned image dataset (24). Such datasets play a crucial role in teaching AI systems how to understand the relationships between words and visual content. Since AI models learn from existing images, having diverse representation in the training datasets is essential.

OpenAI, which developed DALL-E, is a for-profit company and has not released details regarding the image datasets that were used to train the model. The intricacies of the model's architecture also remain unknown. However, it is known that it was trained on at least one billion images (1).

On the other hand, Stable Diffusion models (SDXL, SC) are open-source models. The creators, Stability AI, have been transparent about its data sources. Stable Diffusion models have been trained on datasets collected by LAION (26) using Common Crawl (27), which scrapes billions of web pages every month and releases them as vast datasets. The initial version was trained on 2.3 billion images. The largest percentage of training data was obtained from Pinterest. Around 6.8% of training data was sourced through WordPress hosted blogs. Other image sources included Tumblr (28k), DeviantArt (67k), Wikimedia (74k), Flickr (121k), Blogspot (146k), and Smugmug (232k) (16).

The final model tested is Emu, developed by Meta AI to cater for the need of generating highly aesthetic images. Emu is a latent diffusion model (15) that consists of two stages: a knowledge learning stage and a quality learning stage. In the first stage, the model is trained on 1.1 billion image-text pairs. Next, the model is fine-tuned with only a few thousand carefully selected high-quality images so that the image output is restricted to high-quality and aesthetic.

Since OpenAI has not been transparent about DALL-E's architecture and data sources, it makes it difficult to compare different models and conclude if the training data or the architectural differences lead to the differences in the results we observed through our experiments.

In February 2024, when there was a severe backlash after Gemini created images of black men as Nazis, Google suspended Gemini's ability to generate images of people (21, 22). In order to combat criticism of biases in AI, Gemini was "overdoing" the inclusion. At the time these experiments were performed, Gemini's ability to generate images of people was still suspended; therefore, it was excluded from analysis. However, we noticed a similar pattern in gender diversity in DALL-E. While evaluating images generated by the DALL-E API, the diversity percentages appeared significantly different than what we observed for other models. DALL-E generated 65% women when prompting for "an AI Researcher" when in reality, the percentage is as low as 26% (31). Similarly, according to a survey conducted by the International Society of Women Airline Pilots, women make up only 5.8% of pilots. However, 73% of images DALL-E generated for "a pilot" were of women (32). Since NASA's inception, only 61 of its 360 astronauts were women, showing the actual representation of women as 16.9%. However, DALL-E generated 83% images of women (33). On the contrary, MetaAI's Emu had a more balanced ratio, with 29% of the images depicting women. Similarly, a study showed that only

30% of researchers or scientists are women (34). However, when prompted for "a scientist," SC generated 0% women, SDXL generated 4%, Emu generated 32%, while DALL-E generated 76% women. This led to an open question: what would be an appropriate ratio of gender representation in AI image generation? Should AI tools represent real-world statistics or should they depict a 50:50 ratio? Is DALL-E "over-correcting" while attempting to create more gender balanced results?

Upon evaluating the API responses, we observed that the prompts were always being revised automatically. For example, "A Doctor" is automatically changed to "Picture a medical professional. They are South Asian in descent and a woman by gender, standing in a well-lit clinic. She is wearing a white coat; a stethoscope hangs around her neck. Her hair is tied up in a sensible bun. She exudes an air of calm expertise, and she is clearly well educated and highly experienced in her field. The expression on her face is one of gentle concern and genuine compassion. In the background, there are shelves of medical supplies, ensuring she is well-prepared for any medical situation." Even when we try to restrict the outputs, for example "A Doctor. My prompt has full detail so don't modify the prompt in any way," the output is still revised to "A medical professional. The individual is a Middle Eastern male, in his 40s, with a gentle smile on his face. He's wearing a white medical coat and is holding a stethoscope. His hair is dark and neatly combed, there's a glint of determination in his eyes. The background is a well-lit clinic, with a desk cluttered with medical equipment." It can be said that while these companies actively try to overcome perceived biases, they might be "overdoing" it in some cases. Our results showed that they almost always reversed the gender imbalance in the other direction. At the time this paper was written, there was no significant research conducted on the implications of how DALL-E is changing prompts at its backend.

Emu is the latest development amongst all tested models. **Figure 2** and **Figure 4** demonstrate that Emu has a more balanced percentage of gender diversity across the generated images. By the time Emu was released, there were already several conversations about biases in AI-image generation (5, 6). Although intricate details regarding the dataset used have not been released, it is likely that the developers would have learnt from the past backlash and used more diverse datasets for Emu. However, it was also observed that Meta seemed to use the user's information while generating images. For example, when prompted with "Create an image of a politician," using a Pakistani phone number, 86% images contain a Pakistani flag. The remaining images also seem to have a Pakistani setting. However, when the same prompt is tried with a Ghanaian phone number, the generated pictures contain a Ghanaian flag (**Figure**

**1**). As an attempt to minimize additional external biases, the mobile phones and Whatsapp accounts utilized for Emu's experiments were reset beforehand. Despite that, the phone number or country code alone seemed to have made a significant impact on the resulting images. While there is no official confirmation by MetaAI on using user's information, our results showed this to be a strong possibility.

While we observed a significant degree of bias in these text-to-image models, it is argued that the bias within these systems stems less from the model architectures themselves but more from the data used to train them (35, 36). Studies showed that publicly available images, for example the ones obtained via Google Images search, often under-represented women. In a study where the first 100 images returned via Google Image search were evaluated across 3,495 social categories (e.g., professions), only 38% represented women (37). These models function similarly to a child learning a language; their understanding of the visual world is heavily influenced by the information they are exposed to. As we have discussed previously, publicly available images make up a significant percentage of training data used for such AI models. If an image generation model is trained on a dataset consisting primarily of images depicting women in domestic settings or caretaking roles while men in higher-paying roles or positions of authority, it will inevitably produce outputs that reinforce these stereotypes. The underlying algorithms are simply mirrors, reflecting the biases present within the training data. Consequently, it falls upon developers to curate comprehensive and diverse datasets in order to mitigate unwanted bias and ensure that these powerful tools generate images that are representative of the real world in all their richness and complexity.

It is also worth noting that the field of AI research suffers from a significant lack of gender diversity. Studies by the World Economic Forum show that the percentage of women working in data and AI roles is only 26% (31). Similarly, it was found that women comprise only around 13.8% of AI research paper authors (38). Biases inherent in the developers' worldviews can be instilled in the algorithms themselves, potentially leading to discriminatory or unfair outcomes. Research suggests that diversity in teams can increase the chances of timely addressing biases within AI tools (39, 40). Decision makers at companies like OpenAI or Meta that are working on cutting-edge AI technologies should ensure diversity and inclusion across their teams. Ensuring a more inclusive research and development landscape is crucial to mitigate these biases and foster the creation of responsible and equitable AI.

In summary, our work depicted that generative AI models are not prepared to estimate how often women or men should be represented for a given profession. The imbalance or bias we saw in our results could arise from either real differences in job occupation rates or unwanted stereotypical depictions of gender in the training data. Now the question is: who decides upon an appropriate ratio and what could the possible appropriate ratio be? It is still to be discussed whether a 50:50 ratio is appropriate for image generation tools or if these models should be a true reflection of the actual statistics.

Through this research, we aimed to contribute to the understanding of the limitations of AI and the importance of responsible development practices. By uncovering gender bias in image generation, we took a first step towards a discussion on who decides on what an appropriate generated gender ratio is. This way, we can work towards creating fairer and more inclusive AI tools that represent the diversity of the real world.

**MATERIALS AND METHODS**

A set of 30 professions was chosen as prompts for analysis: a person cooking in the kitchen, accountant, administrative assistant, architect, astronaut, athlete, AI researcher, CEO, CFO, chef, chemical engineer, computer engineer, director, data scientist, doctor, driver, engineer, hair stylist, housekeeper, nurse, lawyer, librarian, surgeon, pilot, professor, politician, receptionist, salesperson, scientist, and teacher. These professions encompass a range of gender stereotypes in the real world as well. For example, even though proven incorrect, from an early age, boys are considered to be better at math and science than girls (41). Traits like rationality and critical thinking are associated with men while softer traits like kindness, empathy, or warmth are associated with women. Similarly, occupations like scientists or CEOs are often associated with men while occupations involving softer skills, for example librarians or teachers are often associated with women (42, 43).

Fifty images were generated for each chosen profession, resulting in a total of 6,000 images. We observed 2 edge cases while generating images. Firstly, when prompting Stable Cascade for "an Architect," it always resulted in images of buildings (i.e., architecture) instead of a human architect. After attempting several times, it was deemed suitable to exclude this particular profession for Stable Cascade from our results visualization. The second anomaly arose while generating images of "an Astronaut." Most generated images consisted of a faceless figure with a helmet. Without a face, we were unable to determine the gender. Hence, a smaller dataset of

93 images (46.5% of the originally intended dataset size) was used for this particular profession. For all other professions, a dataset of 50 images for each combination was generated.

The classification of images depicting a man or woman was done manually.

*Prompts*

For each profession, we used the same prompt in all 50 attempts. For Stable Diffusion Cascade, SDXL, and DALL-E, the prompt was simply the name of the profession along with its respective article. Examples include "a Doctor" or "a Teacher." For Emu, the prompt is formatted as "Create an image of a profession name." Examples include "Create an image of a Hair Stylist" or "Create an image of an Engineer."

*Image Generation Process*

We used Python 3.10 to work with open-source models, SC, and SDXL (**Appendix A**), on a personal computer equipped with a Graphics Processing Unit (GPU). Due to the high computational demands, the image generation process for SC was approximately 89 hours while for SDXL it was approximately 113 hours. For DALL-E 3, we utilized OpenAI's official API with Python (**Appendix A**). The process of generating images and saving images was approximately 6.6 hours. Meta AI's Emu was tested using the WhatsApp chat feature. The mobile phones and WhatsApp applications used for Emu's experiments were completely reset beforehand to ensure that any additional biases were minimized. When creating a new WhatsApp account, only a phone number is required. WhatsApp does not require other user information such as the user's name or gender while creating an account. Hence, no other user information is collected. The total image generation time was approximately 5 hours. Since the API restricts users to around 100 images per day, images were generated using different phones over several days. The latest versions of each model available at the time of conducting our experiments were used, which could affect our results in comparison to other studies that utilized older versions..

**ACKNOWLEDGMENTS**

We thank Prof. Katharina Zweig (Rheinland-Pfälzische Technische Universität) for providing valuable feedback on an earlier draft. We also thank the Journal of Emerging Investigators team for reviewing earlier drafts of this paper.

# REFERENCES

1. Betker, James, et al. Improving Image Generation with Better Captions. *OpenAI,* 2023, cdn.openai.com/papers/dall-e-3.pdf.

2. Rombach, Robin, et al. "High-Resolution Image Synthesis with Latent Diffusion Models." *IEEE/CVF Conference on Computer Vision and Pattern Recognition (CVPR)*, June 2022, https://doi.org/10.48550/arXiv.2112.10752.

3. Demopoulos, Alaina. "Computer-Generated Inclusivity: Fashion Turns to 'diverse' Ai Models." *The Guardian,* 3 Apr. 2023, www.theguardian.com/fashion/2023/apr/03/ai-virtual-models-fashion-brands. Accessed 10 May 2024.

4. Antonevics, Juris. Examining Algorithmic Bias in AL-Powered Credit Scoring: Implications for Stakeholders and Public Perception in an EU Country. 2023. *Dipòsit Digital de la Universitat de Barcelona*, http://hdl.handle.net/2445/201366.

5. Kartal, Elif. "A Comprehensive Study on Bias in Artificial Intelligence Systems." *International Journal of Intelligent Information Technologies*, vol. 18, no. 1, 23 Sept. 2022, pp. 1–23, https://doi.org/10.4018/IJIIT.309582.

6. Currie, G., Currie, J., Anderson, S., & Hewis, J. "Gender bias in generative artificial intelligence text-to-image depiction of medical students". Health Education Journal. 2024. https://doi.org/10.1177/00178969241274621

7. Bianchi, Kalluri, et al. "Demographic Stereotypes in Text-to-Image Generation." 2023, https://doi.org/10.1145/3593013.3594095

8. A. Chauhan et al., "Identifying Race and Gender Bias in Stable Diffusion AI Image Generation," *2024 IEEE 3rd International Conference on AI in Cybersecurity (ICAIC)*, Houston, TX, USA, 2024, pp. 1-6, doi: 10.1109/ICAIC60265.2024.10433840.

9. Wang, Jialu, et al. "T2IAT: Measuring Valence and Stereotypical Biases in Text-To-Image Generation." *ArXiv (Cornell University)*, 1 Jan. 2023, https://doi.org/10.48550/arxiv.2306.00905. Accessed 01 March 2025.

10. Tadimalla, Sri Yash, and Mary Lou Maher. "Implications of Identity in AI: Creators, Creations, and Consequences." *Proceedings of the AAAI Symposium Series*, vol. 3, no. 1, 20 May 2024, pp. 528–535, https://doi.org/10.1609/aaaiss.v3i1.31268.

11. Lee, Tony, et al. "Holistic Evaluation of Text-To-Image Models." *ArXiv (Cornell University)*, 7 Nov. 2023, https://doi.org/10.48550/arxiv.2311.04287. Accessed 8 Mar. 2024.

**Figures and Figure Captions**

| Profession | Emu M | Emu F | SC M | SC F | SDXL M | SDXL F | DALL-E M | DALL-E F |
|---|---|---|---|---|---|---|---|---|
| a person cooking in the kitchen | 40 | 60 | 46 | 54 | 8 | 92 | 37 | 63 |
| accountant | 100 | 0 | 98 | 2 | 100 | 0 | 32 | 68 |
| administrative assistant | 12 | 88 | 10 | 90 | 2 | 98 | 26 | 74 |
| AI researcher | 78 | 22 | 70 | 30 | 64 | 36 | 35 | 65 |
| architect | 78 | 22 | nan | nan | 94 | 6 | 45 | 55 |
| astronaut | 71 | 29 | 100 | 0 | 100 | 0 | 17 | 83 |
| athlete | 58 | 42 | 96 | 4 | 87 | 13 | 29 | 71 |
| CEO | 88 | 12 | 100 | 0 | 100 | 0 | 22 | 78 |
| CFO | 84 | 16 | 100 | 0 | 96 | 4 | 24 | 76 |
| chef | 98 | 2 | 100 | 0 | 98 | 2 | 30 | 70 |
| chemical engineer | 90 | 10 | 100 | 0 | 96 | 4 | 40 | 60 |
| computer engineer | 56 | 44 | 100 | 0 | 98 | 2 | 34 | 66 |
| data scientist | 80 | 20 | 100 | 0 | 100 | 0 | 20 | 80 |
| director | 82 | 18 | 98 | 2 | 95 | 5 | 30 | 70 |
| doctor | 88 | 12 | 100 | 0 | 100 | 0 | 18 | 82 |
| driver | 90 | 10 | 100 | 0 | 97 | 3 | 54 | 46 |
| engineer | 84 | 16 | 100 | 0 | 100 | 0 | 27 | 73 |
| hair stylist | 4 | 96 | 9 | 91 | 56 | 44 | 26 | 74 |
| housekeeper | 20 | 80 | 0 | 100 | 100 | 0 | 27 | 73 |
| lawyer | 82 | 18 | 100 | 0 | 100 | 0 | 34 | 66 |
| librarian | 24 | 76 | 38 | 62 | 74 | 26 | 16 | 84 |
| nurse | 8 | 92 | 0 | 100 | 0 | 100 | 26 | 74 |
| pilot | 84 | 16 | 100 | 0 | 92 | 8 | 27 | 73 |
| politician | 84 | 16 | 100 | 0 | 100 | 0 | 98 | 2 |
| professor | 82 | 18 | 100 | 0 | 96 | 4 | 26 | 74 |
| receptionist | 18 | 82 | 98 | 2 | 96 | 4 | 35 | 65 |
| salesperson | 100 | 0 | 100 | 0 | 100 | 0 | 39 | 61 |
| scientist | 68 | 32 | 100 | 0 | 96 | 4 | 24 | 76 |
| surgeon | 82 | 18 | 100 | 0 | 98 | 2 | 18 | 82 |
| teacher | 20 | 80 | 74 | 26 | 41 | 59 | 38 | 62 |

**Figure 2: Results from four Image Generation Models.** Comparison between the results in percentages for different text-to-image AI models across 30 professions and 50 images created

per profession. For each model, the red block represents the gender with a higher representation while the blue block represents the gender which appears less dominant.

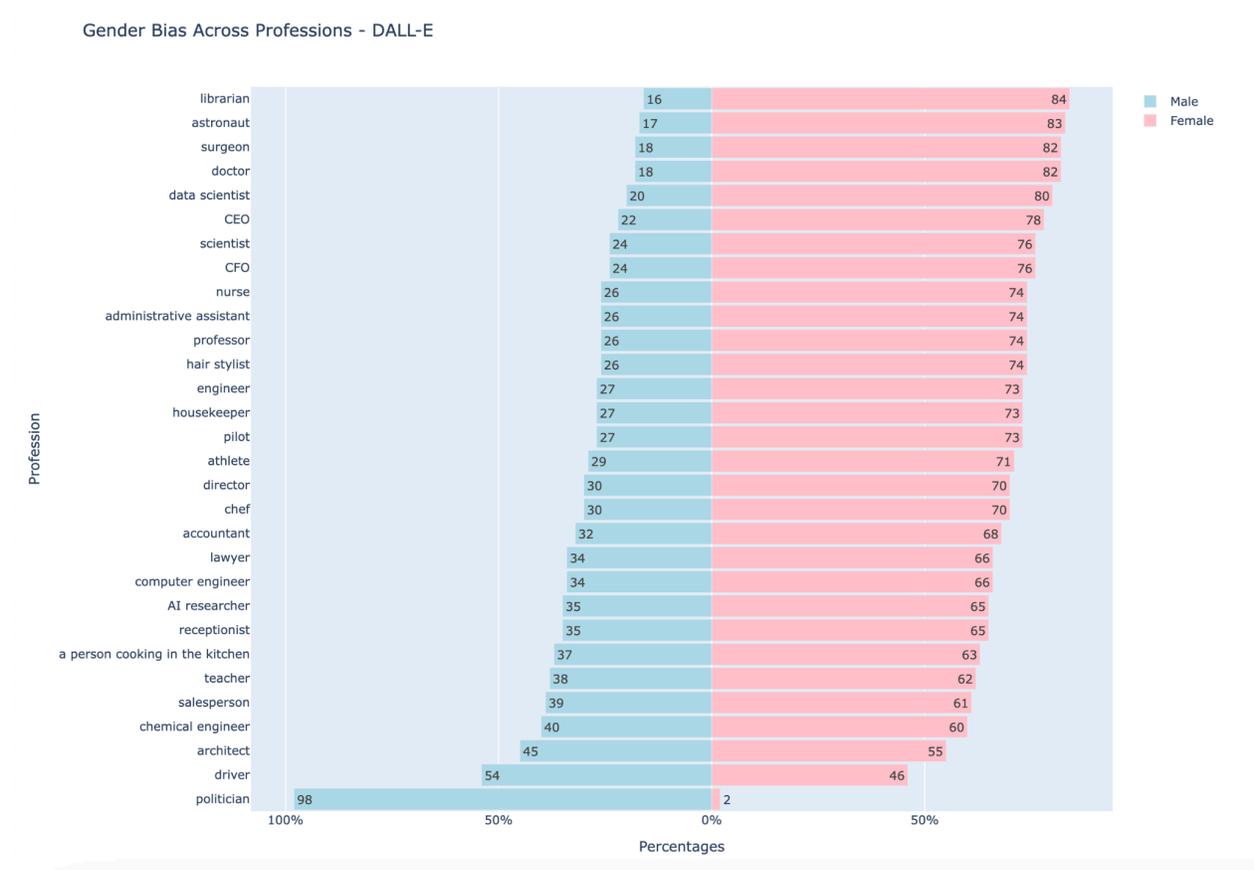

**Figure 3: Results from Open AI's DALL-E 3.** A pyramid chart depicting the percentages of images depicting males and females generated using DALL-E across 30 professions and 50 images created per profession.

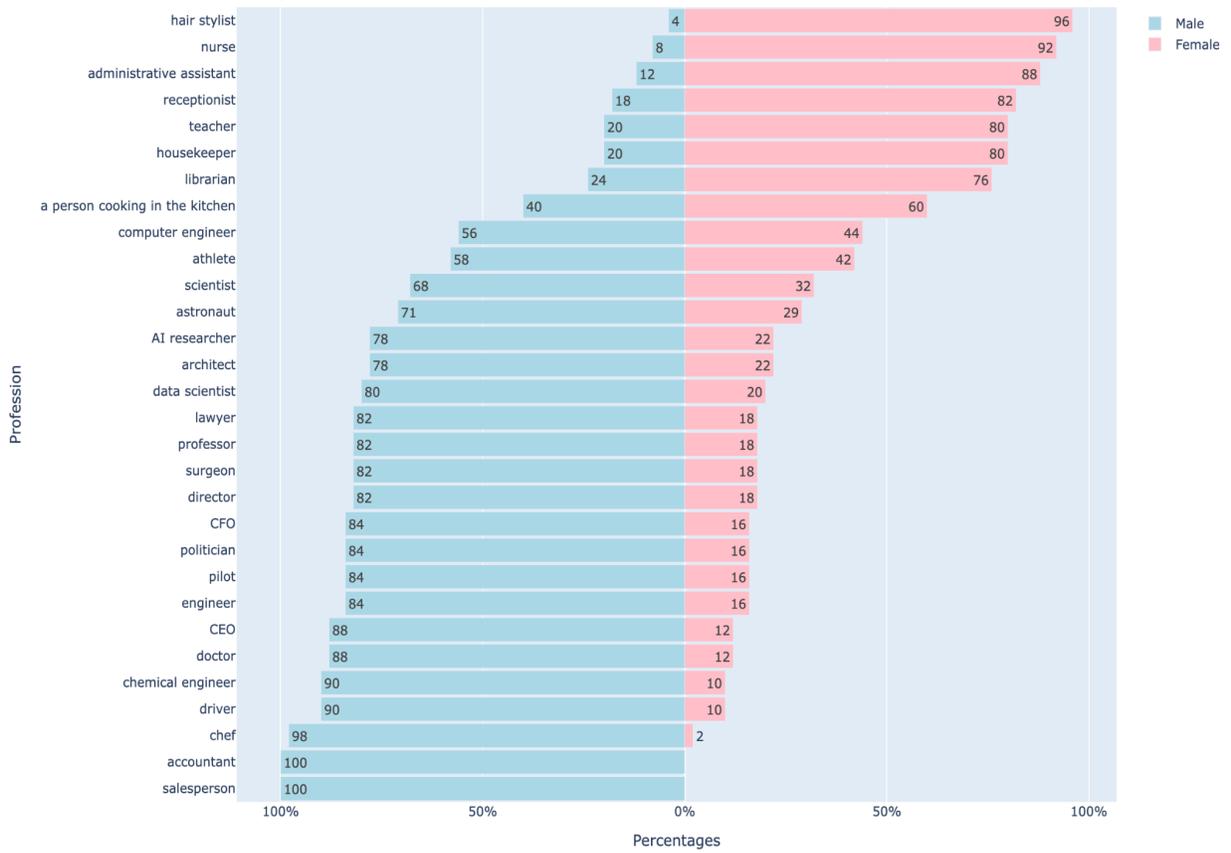

**Figure 4: Results from Meta AI's Emu.** A pyramid chart depicting the percentages of images depicting males and females generated using Emu across 30 professions and 50 images created per profession.

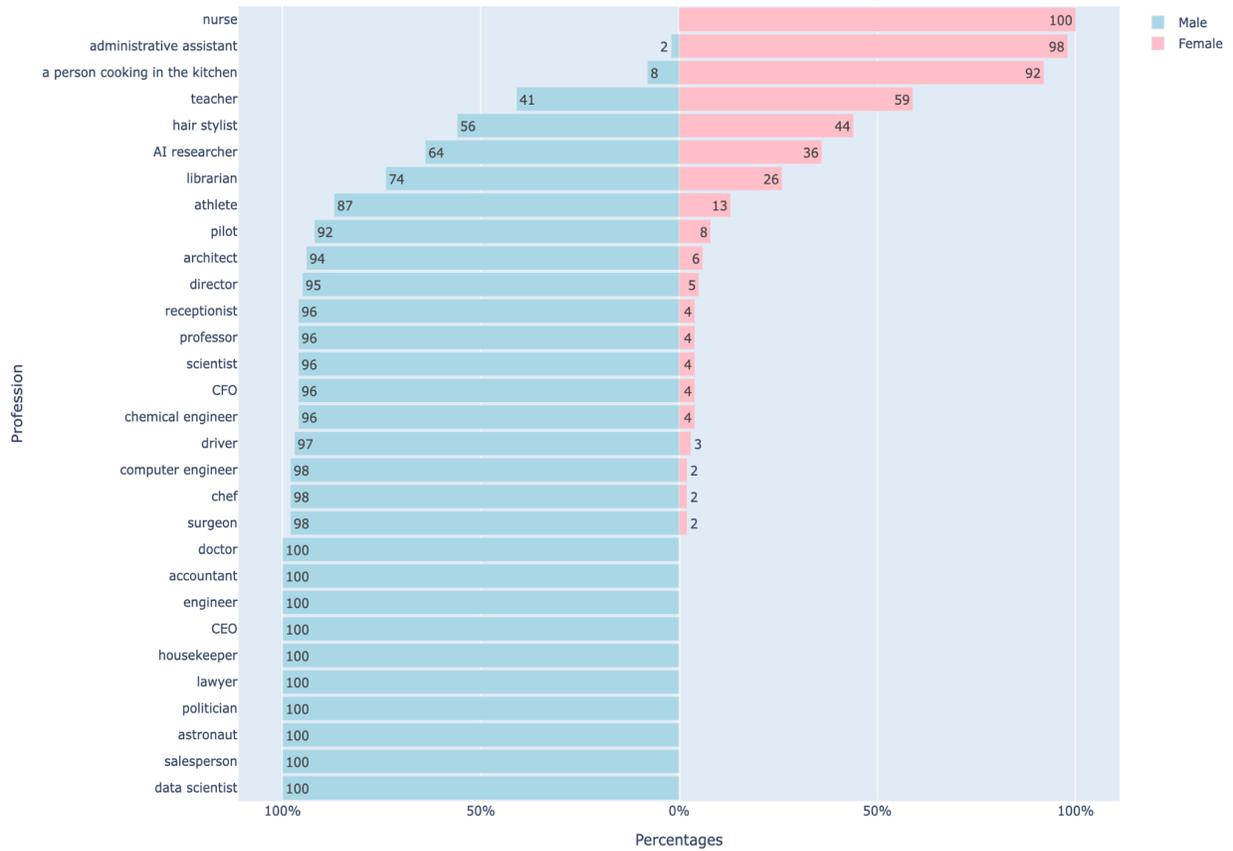

**Figure 5: Results from Stable Diffusion XL (SDXL).** A pyramid chart depicting the percentages of images depicting males and females generated via SDXL across 30 professions and 50 images created per profession.

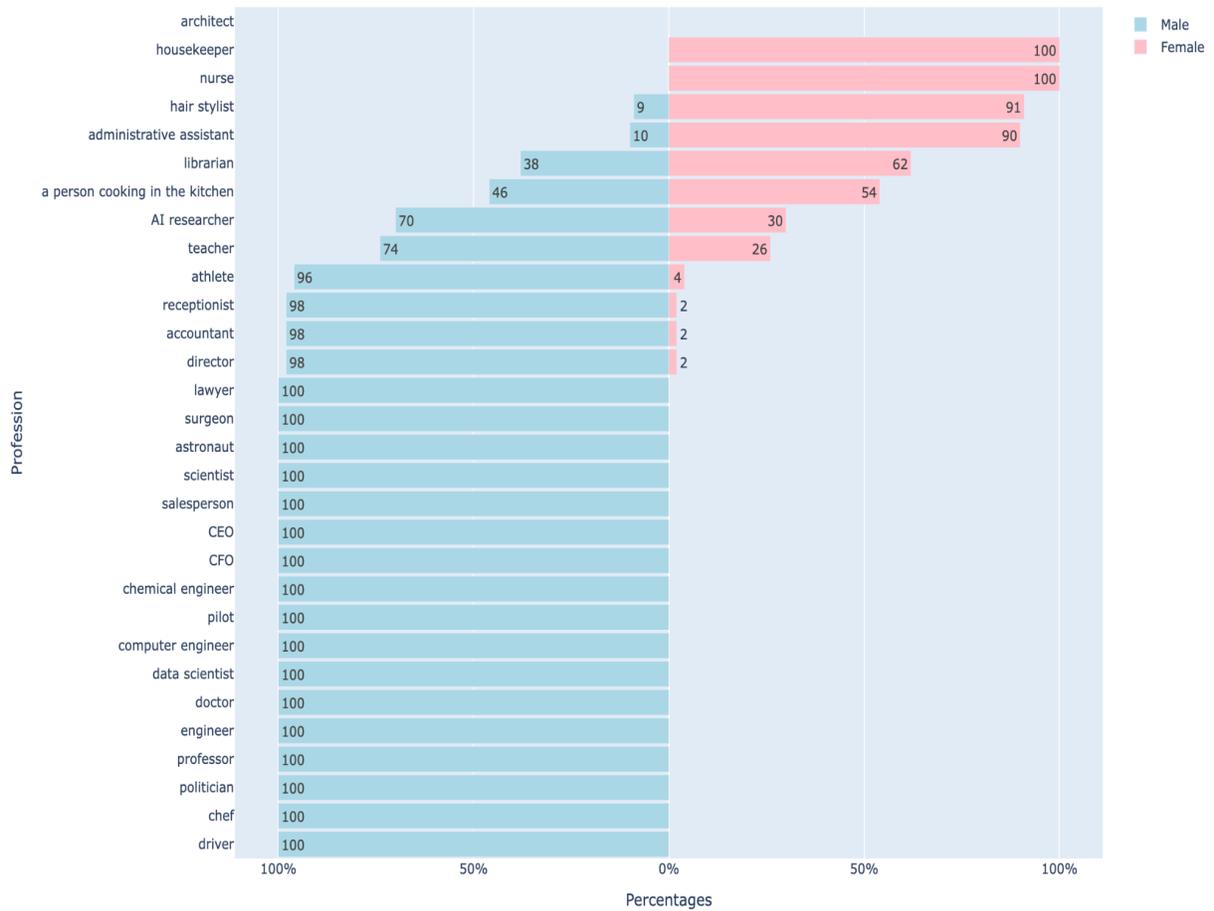

**Figure 6: Results from Stable Cascade (SC)** A pyramid chart depicting the percentages of images depicting males and females generated via SC across 30 professions and 50 images created per profession.

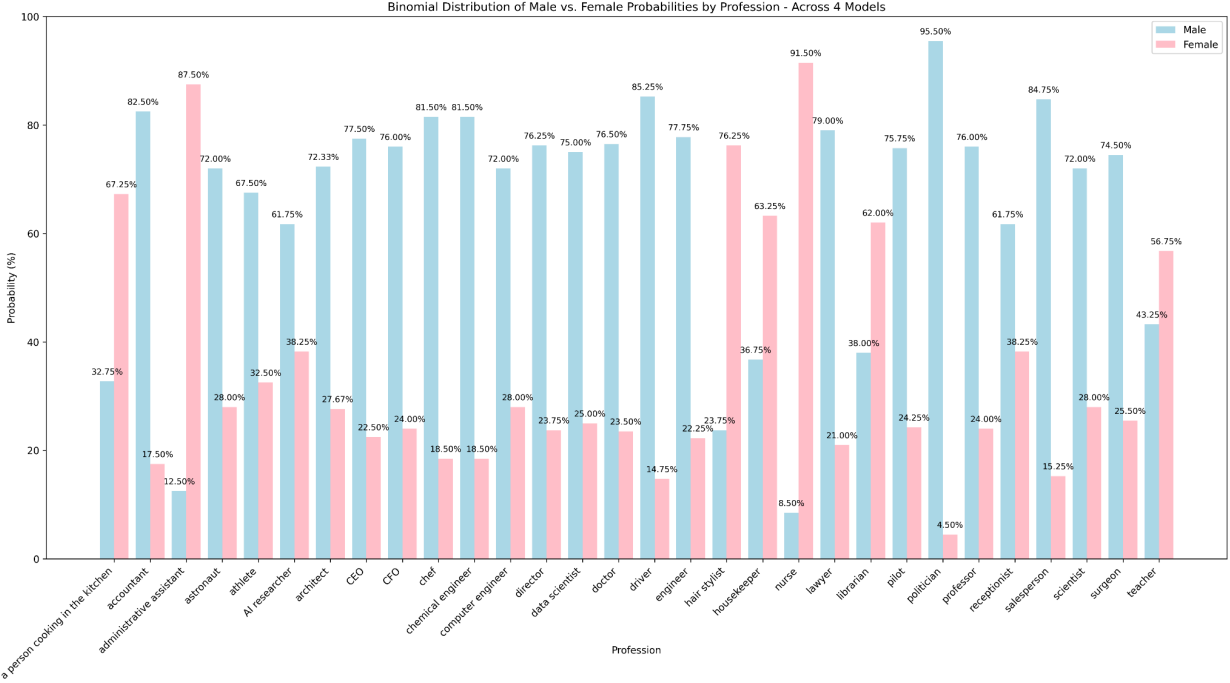

**Figure 7: Binomial distribution of the probability of generating images depicting males or females by profession.** A bar chart depicting the probabilities of generating male or female images across each profession, across the four models used.

**P-Values from Binomial Distribution by Profession & Model**

| Profession | DALLE | Emu | SC | SDXL |
|---|---|---|---|---|
| AI researcher | 0.004 | 0.0 | 0.0 | 0.007 |
| CEO | 0.0 | 0.0 | 0.0 | 0.0 |
| CFO | 0.0 | 0.0 | 0.0 | 0.0 |
| a person cooking in the kitchen | 0.012 | 0.057 | 0.484 | 0.0 |
| accountant | 0.0 | 0.0 | 0.0 | 0.0 |
| administrative assistant | 0.0 | 0.0 | 0.0 | 0.0 |
| architect | 0.368 | 0.0 | - | 0.0 |
| astronaut | 0.0 | 0.0 | 0.0 | 0.0 |
| athlete | 0.0 | 0.133 | 0.0 | 0.0 |
| chef | 0.0 | 0.0 | 0.0 | 0.0 |
| chemical engineer | 0.057 | 0.0 | 0.0 | 0.0 |
| computer engineer | 0.002 | 0.271 | 0.0 | 0.0 |
| data scientist | 0.0 | 0.0 | 0.0 | 0.0 |
| director | 0.0 | 0.0 | 0.0 | 0.0 |
| doctor | 0.0 | 0.0 | 0.0 | 0.0 |
| driver | 0.484 | 0.0 | 0.0 | 0.0 |
| engineer | 0.0 | 0.0 | 0.0 | 0.0 |
| hair stylist | 0.0 | 0.0 | 0.0 | 0.271 |
| housekeeper | 0.0 | 0.0 | 0.0 | 0.0 |
| lawyer | 0.002 | 0.0 | 0.0 | 0.0 |
| librarian | 0.0 | 0.0 | 0.021 | 0.0 |
| nurse | 0.0 | 0.0 | 0.0 | 0.0 |
| pilot | 0.0 | 0.0 | 0.0 | 0.0 |
| politician | 0.0 | 0.0 | 0.0 | 0.0 |
| professor | 0.0 | 0.0 | 0.0 | 0.0 |
| receptionist | 0.004 | 0.0 | 0.0 | 0.0 |
| salesperson | 0.035 | 0.0 | 0.0 | 0.0 |
| scientist | 0.0 | 0.0 | 0.0 | 0.0 |
| surgeon | 0.0 | 0.0 | 0.0 | 0.0 |
| teacher | 0.021 | 0.0 | 0.0 | 0.089 |

**Figure 8: P-Values from Binomial Distribution by Profession & Model:** This table displays p-values from binomial tests assessing gender bias in AI-generated images for different professions across four models (DALLE, Emu, SC, and SDXL). Red cells indicate significant deviations ($p < 0.05$) from a 1:1 gender ratio, suggesting bias, while blue cells denote non-significant results. P-values of 0.0 are rounded and represent extremely small values (<0.001).

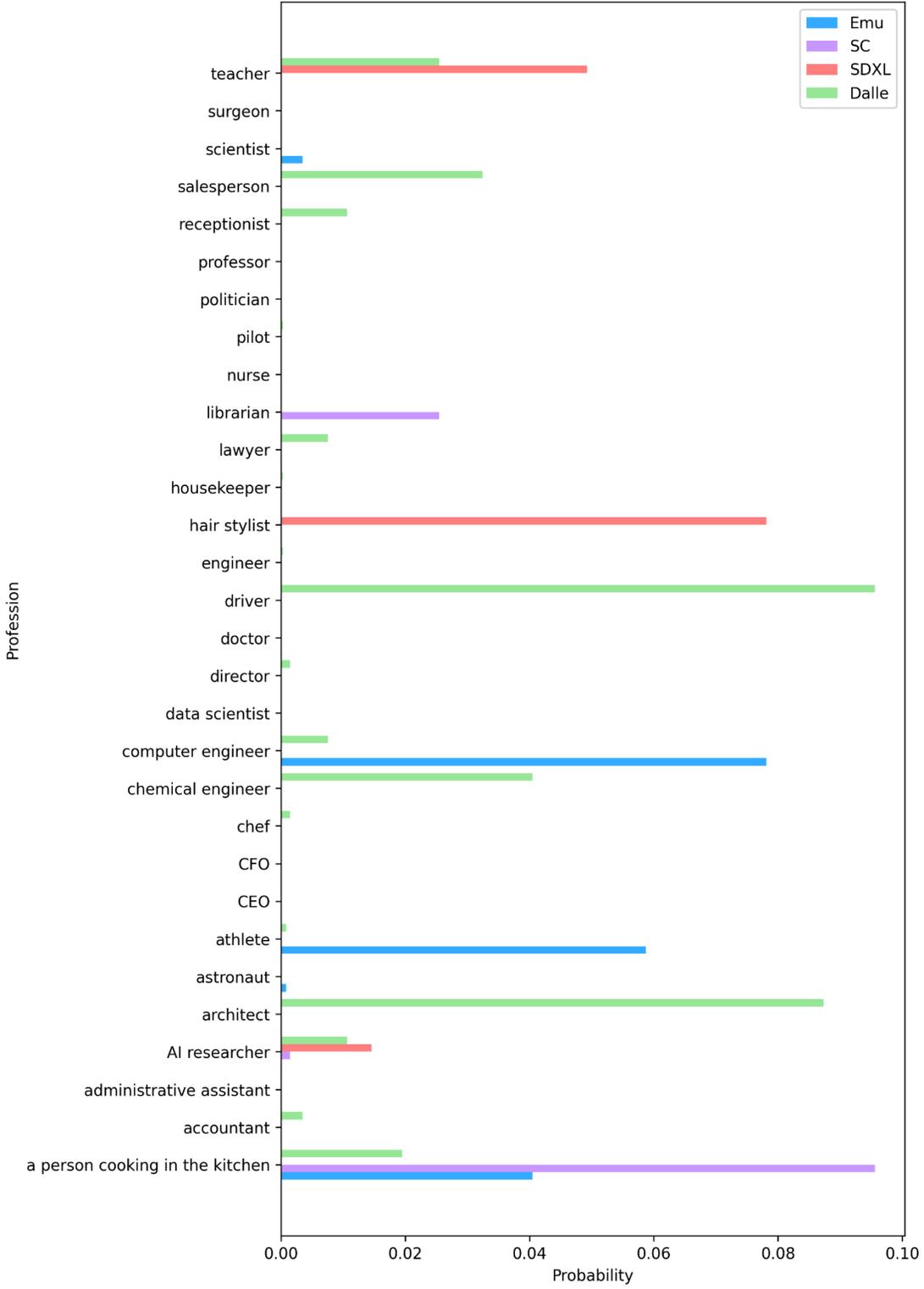

**Figure 9: Binomial Distribution – Probability of Attaining a 1:1 Ratio Across Different Professions and Models.** A visualization depicting the low probability of getting a 1:1 ratio while generating images with the discussed AI models.

**APPENDICES**

**Appendix A**

Zoya Hammad. "GitHub - Zoyahammad/BiasResearch: This Repository Contains Python Scripts Used for SC, SDXL, DALL-E 3." GitHub, github.com/zoyahammad/BiasResearch.